\documentclass[aps,twocolumn,longbibliography,superscriptaddress]{revtex4-1}
\pdfoutput=1

\usepackage{amsmath,amssymb,mathtools,bm,color,graphicx,empheq,lipsum}
\usepackage[colorlinks=true, allcolors=blue]{hyperref}

\newcommand{\mat}[1]{\begin{pmatrix}#1\end{pmatrix}}

\begin{document}

\title{Searching for Bloch wave packets with almost definite momentum direction}

\author{Arseni Goussev}
\affiliation{Section of Mathematics, University of Geneva, Rue du Conseil-G\'en\'eral 7-9, 1205 Gen\`eve, Switzerland}
\affiliation{Quantum Physics Corner Ltd, 20--22 Wenlock Road, London N1 7GU, United Kingdom}

\author{Gregory V. Morozov}
\affiliation{Scottish Universities Physics Alliance (SUPA), School of Computing, Engineering and Physical Sciences, University of the West of Scotland, Paisley PA1 2BE, United Kingdom}

\date{\today}

\begin{abstract}
	The motion of a quantum particle in a one-dimensional periodic potential can be described in terms of Bloch wave packets. Like free-particle wave packets, they can propagate without attenuation. Here, we examine this similarity more closely by investigating whether Bloch wave packets can maintain a definite -- or nearly definite -- momentum direction, a property inherent to free-particle wave packets. This question is particularly relevant to the feasibility of using solid-state-based systems in the search for the first experimental realization of quantum backflow, a quantum interference effect in which a particle's probability density flows in a direction opposite to its momentum.
\end{abstract}

\maketitle

\section{Introduction}

When a quantum particle moves through a periodic potential, its behavior differs fundamentally from that of a classical particle. According to the celebrated Bloch theory, a perfectly periodic potential becomes effectively transparent to a quantum matter wave of suitable energy, allowing it to propagate ``without attenuation because of the coherent constructive interference of the scattered waves'' \cite{Ashcroft_Mermin}. As Kittel states it, ``matter waves can propagate freely in a periodic structure,'' and ``Bloch functions can be assembled into wave packets to represent electrons that propagate freely through the potential field of the ion cores'' \cite{Kittel}. This free-particle-like motion of an electron through the periodic potential of a crystal is crucial for understanding the electrical properties of solids, such as the emergence of ballistic transport in mesoscopic systems \cite{Hol99Quantum, Jal16Mesoscopic}. The resemblance between quantum particle propagation in periodic potentials and in free space suggests that mesoscopic systems could be used to explore quantum backflow -- a classically impossible effect in which the probability density of a particle moves in a direction opposite to its momentum \footnote{The body of literature on quantum backflow is substantial, and a full review is beyond our scope. Some of the key theoretical developments can be found in Refs.~\cite{BM94Probability, EFV05Quantum, Ber10Quantum}. For an accessible introduction, we refer the reader to Ref.~\cite{YH13introduction}, while Refs.~\cite{Bra21Probability, BG25Quantum} provide a comprehensive list of more recent advancements in the field.}.

Quantum backflow is typically considered for a free particle on a line. In the original formulation of the effect, the particle is prepared in a quantum state with a definite momentum direction, assumed to be positive for concreteness. Since free-space evolution conserves the momentum direction, any measurement of the particle's momentum at a later time is guaranteed to yield a positive value, albeit of a random magnitude. Quantum backflow then manifests as the emergence of a negative probability current at certain points in space and time, meaning that despite having positive momentum, the particle can locally move in the negative direction.

The physical mechanism behind quantum backflow rests on the idea that interference among positive-momentum components within a quantum state can produce a negative probability current. In principle, the effect can occur not only in states composed exclusively of positive-momentum components but also in those containing a small fraction of negative momentum, provided their contribution remains negligible compared to the interference-induced negative flux. This reasoning has enabled researchers to investigate quantum backflow in superpositions of Gaussian wave packets that include a minor admixture of negative momentum components~\cite{YHHW12Analytical, PTMM13Detecting, MPM+14Interference}.

A promising approach to exploring quantum backflow in states with arbitrary momentum distributions was proposed in Ref.~\cite{MYDP21Experiment}, introducing an alternative definition of the effect based on a finite-precision joint probability density function for a particle’s position and momentum. This formulation allows one to determine the threshold of negative probability current compatible with classical mechanical evolution. Any excess of negative current beyond this threshold is then attributed to quantum backflow. However, it has been shown that this new criterion is highly sensitive to the choice of the phase-space probability density function and can, in some cases, yield results inconsistent with the original definition of quantum backflow for positive-momentum states \cite{BG21experiment}.

Although quantum backflow was first recognized as a distinct phenomenon over three decades ago \cite{BM94Probability}, it has yet to be experimentally demonstrated. Significant progress has been achieved in laboratory studies of its optical analog \cite{EZB20Observation, DGGL22Demonstrating, GDGL23Azimuthal, ZHD+25Observation}, but no experiment has successfully observed quantum backflow with massive particles, such as electrons or atoms. (An experimental scheme utilizing Bose-Einstein condensates was proposed in Refs.~\cite{PTMM13Detecting, MPM+14Interference}, but to our knowledge, it has never been realized in a laboratory.) The challenge lies in the fact that quantum backflow is inherently weak -- according to theory, less than $4\%$ of the total probability can be transported against the particle’s momentum \cite{BM94Probability, EFV05Quantum, PGKW06new} -- as well as in the exotic nature of the theoretical backflow-maximizing state, which is spatially unbounded and has infinite mean energy \cite{YHHW12Analytical}. (Some of these difficulties can be mitigated by considering quantum backflow in a circular geometry \cite{Gou21Quantum}.) Identifying favorable physical systems is therefore crucial for making the first experimental observation of the effect possible, as well as for exploring potential technological applications.

A compelling strategy to realizing quantum backflow experimentally is to use electrically charged particles moving along an effectively one-dimensional channel, as proposed in Ref.~\cite{BM94Probability}. Since the electric current is proportional to the probability current, detecting a current (divided by charge) flowing opposite to the particle's momentum would serve as experimental confirmation of quantum backflow. In this regard, an electronic wave packet propagating ballistically through a mesoscopic quantum wire presents an appealing experimental scenario. (Ref.~\cite{BGM+18Coherent} reviews various experimental methods for the generation and manipulation of single-electron wave packets in mesoscopic systems.) This raises a crucial question: does the dynamics of an electron in the periodic potential of a crystal lattice allow for time-dependent states with positive momentum, which are desirable for accessing quantum backflow physics? In other words, do there exist normalizable superpositions of Bloch waves \footnote{For any periodic potential, the set of all Bloch waves forms a complete basis \cite{Cot71Floquet}. This implies that every normalized wave packet capable of propagating through the periodic potential can be represented as a superposition of Bloch waves.} -- henceforth referred to as Bloch wave packets -- whose momentum maintains a definite direction at all times?  If not, what is the highest possible probability that a momentum measurement performed on a Bloch wave packet yields a positive value? Moreover, does this probability remain constant, as in the free-particle case, or does it change in time as the particle moves through the crystal? These questions are the focus of the present study.

While the properties of stationary Bloch waves in one-dimensional periodic potentials have been extensively studied \cite{Cot71Floquet, EasthamFloquet, Morozov2011Floquet, Vega2021Floquet}, time-dependent Bloch wave packets have received relatively little attention. Nonetheless, some important differences between the motion of Bloch wave packets and that of free particles have been investigated. Kroemer first observed that the propagation of a single-band Bloch wave packet with a narrow quasi-momentum distribution is not uniform but rather ``similar to the motion of a caterpillar'': the wave packet undergoes periodic shape variations superimposed on an average velocity, equal to the group velocity \cite{Kro75group}. The origin of these shape variations was later elucidated by Churchill, who showed that a narrow-quasi-momentum wave packet can be expressed as a discrete superposition of ``subpackets'' with distinct but constant shapes, all propagating at the same group velocity while having different phase velocities \cite{Chu77periodic}. Additional insights and methods for analyzing the instantaneous shape, position, and velocity of various features of Bloch wave packet envelopes are discussed in Refs.~\cite{Chu77Instantaneous, CH87Motion}.

In this paper, we identify and explore another key difference between Bloch wave packets and free-particle wave packets: unlike the latter, the former cannot maintain a definite momentum direction. Specifically, we show that Bloch wave packets composed of Bloch waves from a single allowed energy band can never have a definite momentum direction, even instantaneously. In contrast, those involving multiple energy bands may attain a definite direction at certain times but cannot maintain it indefinitely. Then, considering the case of a cosine potential, we examine how its amplitude affects the maximum probability of the particle's momentum being positive. Overall, our findings suggest that in studying quantum backflow in systems with periodic potentials -- such as an electron in a ballistic quantum wire -- it is essential to adopt a formulation that permits backflowing states to include an admixture of negative momentum components. For instance, the state may be constructed so that the fraction of negative momentum is small compared to the interference-induced negative probability transfer, or a phase-space-based formulation akin to that examined in Refs.~\cite{MYDP21Experiment, BG21experiment} may be employed. We further argue that single-band, as opposed to multiple-band, Bloch wave packets are better suited for studies of quantum backflow, as they more closely resemble free-particle wave packets: the fraction of negative momentum components remains constant during time evolution, rather than varying over time.

The paper is organized as follows. Section~\ref{sec:preliminaries} presents an intuitive argument suggesting that a one-dimensional quantum particle in a non-constant potential is unlikely to maintain a definite momentum direction. A more rigorous analysis follows in Sections~\ref{sec:Bloch_WPs} and \ref{sec:p>0}. In particular, Section~\ref{sec:Bloch_WPs} introduces the concept of a Bloch wave packet and establishes the notation used throughout the paper. Section~\ref{sec:p>0} examines the probability that a particle in a periodic potential has positive momentum. We demonstrate that when a wave packet consists solely of Bloch waves from the same energy band, this probability remains constant but strictly less than one. However, if the wave packet contains contributions from different energy bands, the probability oscillates over time, with its time average still falling below one. Section~\ref{sec:V=cos} explores a specific case in which the periodic potential has a cosinusoidal form. We determine how the least upper bound on the probability of positive momentum depends on the potential amplitude and derive analytical approximations for this dependence in the weak- and strong-potential regimes. Finally, Section~\ref{sec:conclusion} summarizes our findings and discusses their implications in the context of the quantum backflow problem.

\section{Preliminary considerations}
\label{sec:preliminaries}

We begin by presenting some preliminary considerations to clarify what it means for the momentum of a one-dimensional particle to have a definite direction, which, for concreteness, we take to be positive. Using Dirac notation, we denote the particle's state at time $t = 0$ by $| \Psi_0 \rangle$, assuming it is normalized to unity, $\langle \Psi_0 | \Psi_0 \rangle = 1$. We then introduce the projection operator onto the subspace of positive momenta,
\begin{equation}
	\hat{\mathcal{P}}_+ = \int_0^{\infty} dp \, | p \rangle \langle p | \,,
\end{equation}
where $| p \rangle$ represents the eigenstate of the momentum operator corresponding to the eigenvalue $p$. At $t = 0$, the particle's momentum is positive with certainty (and thus has a definite direction) if and only if
\begin{equation}
	\hat{\mathcal{P}}_+ | \Psi_0 \rangle = | \Psi_0 \rangle \,.
\label{p>0_condition_1}
\end{equation}
Clearly, any state of the form $| \Psi_0 \rangle = \int_0^{\infty} dp \, \varphi(p) | p \rangle$, where $\varphi(p)$ is a complex-valued function, satisfying $\int_0^{\infty} dp \, |\varphi(p)|^2 = 1$ to ensure normalization, fulfills Eq.~\eqref{p>0_condition_1}  and represents a positive-momentum particle. It is therefore always possible to construct a quantum state with positive momentum at a given instant.

In order for the particle's momentum to remain positive (with certainty) throughout its time evolution, $| \Psi_0 \rangle$ has to satisfy
\begin{equation}
	\hat{\mathcal{P}}_+ \, \hat{\mathcal{U}}(t) | \Psi_0 \rangle = \hat{\mathcal{U}}(t) | \Psi_0 \rangle
\end{equation}
for all $t$, where $\hat{\mathcal{U}}(t)$ is the evolution operator. Differentiating both sides of the last equation $n$ times with respect to $t$, applying the Schr\"odinger equation $\frac{d}{dt} \hat{\mathcal{U}}(t) = -\frac{i}{\hbar} \hat{\mathcal{H}} \hat{\mathcal{U}}(t)$, where $\hat{\mathcal{H}}$ is the system's Hamiltonian, and evaluating the result at $t = 0$, we obtain the following set of conditions:
\begin{equation}
	\hat{\mathcal{P}}_+ \hat{\mathcal{H}}^n | \Psi_0 \rangle = \hat{\mathcal{H}}^n | \Psi_0 \rangle \,, \quad n \ge 1 \,.
\label{p>0_intermediate_condition}
\end{equation}
Here, the Hamiltonian $\hat{\mathcal{H}}$ is assumed to be time-independent. Then, using Eq.~\eqref{p>0_condition_1}, we replace $| \Psi_0 \rangle$ in the right-hand side of Eq.~\eqref{p>0_intermediate_condition} by $\hat{\mathcal{P}}_+ | \Psi_0 \rangle$ to obtain
\begin{equation}
	[ \hat{\mathcal{P}}_+ , \hat{\mathcal{H}}^n ] | \Psi_0 \rangle = 0 \,, \quad n \ge 1 \,,
	\label{p>0_condition_2}
\end{equation}
where $[\cdot, \cdot]$ denotes the commutator.

Any quantum state $| \Psi_0 \rangle$ representing a particle with positive momentum at all times must satisfy Eqs.~\eqref{p>0_condition_1} and \eqref{p>0_condition_2}. For a free particle, the infinite set of equations \eqref{p>0_condition_2} is trivially fulfilled by any state since the Hamiltonian $\hat{\mathcal{H}}$ consists solely of the kinetic term, which commutes with the projection operator $\hat{\mathcal{P}}_+$. As a result, any free-particle state that has positive momentum at $t = 0$ will retain this property throughout its time evolution. However, for a particle in a periodic potential, the commutators $[ \hat{\mathcal{P}}_+ , \hat{\mathcal{H}}^n ]$ are generally nonzero. This implies that for $| \Psi_0 \rangle$ to represent a state with positive momentum at all times, it must belong to the intersection of the null spaces of infinitely many operators: $(\hat{\mathcal{P}}_+ - 1)$, $[ \hat{\mathcal{P}}_+ , \hat{\mathcal{H}} ]$, $[ \hat{\mathcal{P}}_+ , \hat{\mathcal{H}}^2 ]$, $[ \hat{\mathcal{P}}_+ , \hat{\mathcal{H}}^3 ]$, $\ldots$. Since these operators are generally distinct, it is reasonable to expect that their null spaces have an empty intersection, implying that the desired positive-momentum state $| \Psi_0 \rangle$ does not exist.

This simple argument provides an intuitive basis for expecting the non-existence of states with a definite momentum direction in a general periodic potential, but it does not constitute a proof. More importantly, it offers no insight into how close the probability of finding the particle with positive momentum can be to one. In the following sections, we undertake a detailed analysis of these questions.

\section{Bloch wave packets}
\label{sec:Bloch_WPs}

To set the stage for our study, we first define the concept of a Bloch wave packet and establish the notation used throughout the paper.

We consider a quantum particle of mass $\mu$ in a one-dimensional periodic potential $V(x)$ with period $L$, i.e.~$V(x+L) = V(x)$. Throughout the paper, we assume that the periodic potential has a finite -- though possibly very large -- number of nonzero Fourier coefficients. More concretely, we take
\begin{equation}
	V(x) = \sum_{n=-N}^{N} V_n e^{i n q x} \,,
\label{V(x)}
\end{equation}
where \begin{equation}
	q = \frac{2 \pi}{L}
\end{equation}
is the period of the corresponding reciprocal lattice. Here, the Fourier coefficients $V_n$ satisfy $V_{-n} = V_n^*$ to ensure that $V(x)$ is a real-valued function, and $N$ is such that $V_{-N} = V_N^* \not= 0$. The form given by Eq.~\eqref{V(x)} is quite general and encompasses a wide range of physically meaningful periodic potentials.

The time-independent Schr\"odinger equation, that determines stationary states $F^{(j)}(x,\kappa)$ and energies $E^{(j)}(\kappa)$ of the particle, reads
\begin{equation}
	\left( -\frac{\hbar^2}{2 \mu} \frac{\partial^2}{\partial x^2} + V(x) \right) F^{(j)}(x, \kappa) = E^{(j)}(\kappa) F^{(j)}(x, \kappa) \,.
\label{TISE}
\end{equation}
According to the Bloch theorem \cite{Ashcroft_Mermin, Kittel, Cot71Floquet}, the stationary states -- known as (Floquet-)Bloch waves -- can be expressed as the product of a plane wave $e^{i \kappa x}$ and an $L$-periodic function of $x$, i.e.
\begin{equation}
	F^{(j)}(x, \kappa) = e^{i \kappa x} \sum_{n=-\infty}^{+\infty} f_n^{(j)}(\kappa) e^{i n q x} \,.
\label{Bloch_theorem}
\end{equation}
Here, $f_n^{(j)}(\kappa)$ are the Fourier expansion coefficients of the periodic part of the Bloch wave, while $\hbar \kappa$ represents the quasi-momentum, which, without loss of generality, can be restricted to the first Brillouin zone, $-q/2 < \kappa \le q/2$. The superscript $j = 0, 1, 2, \ldots$ labels the allowed energy bands in order of increasing energy.

The substitution of Eq.~\eqref{Bloch_theorem} into Eq.~\eqref{TISE} gives rise to the following eigenproblem:
\begin{multline}
	\sum_{m = -\infty}^{+\infty} \left( \frac{\hbar^2}{2 \mu} (\kappa + n q)^2 \delta_{m,n} + V_{n-m} \right) f_m^{(j)}(\kappa) \\ = E^{(j)}(\kappa) f_n^{(j)}(\kappa) \,.
\label{F_def}
\end{multline}
Solving this eigenproblem for every quasi-momentum $\hbar \kappa$ on the interval $-q/2 < \kappa \le q/2$, one obtains the eigenvalues $E^{(0)}(\kappa) < E^{(1)}(\kappa) < \ldots$ \footnote{Throughout this paper, we assume that all allowed energy bands are separated by nonzero band gaps. That is, we do not consider exotic periodic potentials where certain band gaps vanish at specific quasi-momentum values.}, representing the allowed energy bands, along with the corresponding eigenvectors $\{ f_n^{(0)}(\kappa) \}_{n=-\infty}^{+\infty}, \, \{ f_n^{(1)}(\kappa) \}_{n=-\infty}^{+\infty}, \, \ldots$, which determine the Bloch waves $F^{(0)}(x, \kappa), \, F^{(1)}(x, \kappa), \, \ldots$. We normalize the eigenvectors according to
\begin{equation}
	\sum_{n = -\infty}^{+\infty} \left| f_n^{(j)}(\kappa) \right|^2 = \frac{1}{2 \pi} \,,
\label{f_normalization}
\end{equation}
for all $\kappa$ and $j$. This choice of the normalization condition for the Fourier coefficients entails the following normalization condition for the Bloch waves (see Appendix~\ref{app:F_normalization} for a proof):
\begin{equation}
	\int_{-\infty}^{+\infty} dx \, F^{(j)}(x,\kappa)^* F^{(j')}(x,\kappa') = \delta(\kappa - \kappa') \delta_{j,j'} \,.
\label{F_normalization}
\end{equation}

A Bloch wave packet $\Psi(x,t)$ is a normalizable superposition of Bloch waves corresponding to different quasi-momenta and energy bands. In general, 
\begin{equation}
	\Psi(x,t) = \sum_{j=0}^{\infty} \int_{-q/2}^{q/2} d\kappa \, \phi^{(j)}(\kappa) e^{-i E^{(j)}(\kappa) t / \hbar} F^{(j)}(x, \kappa) \,,
\label{Bloch_WP}
\end{equation}
where $\phi^{(j)}(\kappa)$ is a generally complex-valued function, representing the quasi-momentum amplitude in the $j^{\text{th}}$ energy band. Wave packets of this form satisfy the time-dependent Schr\"odinger equation, $i \hbar \frac{\partial \Psi}{\partial t} = -\frac{\hbar^2}{2 \mu} \frac{\partial^2 \Psi}{\partial x^2} + V(x) \Psi$, and represent quantum states of a moving particle. In this study, we assume that every amplitude function $\phi^{(j)}(\kappa)$ has its support in the open interval $-q/2 < \kappa < q/2$. In other words, we do not consider wave packets involving Bloch waves $F^{(j)}(x, q/2)$ coming from the Brillouin zone boundary \footnote{At the zone boundary, not all solutions to the time-independent Schr{\"o}dinger equation have the Bloch wave structure \cite{Allen1953IncipientBands, Cot71Floquet, Cot72Solutions, Loly1992IncipientBands, Morozov2018IncipientBands}, and the inclusion of the zone boundary in the support of the amplitudes $\phi^{(j)}(\kappa)$ may require modifying the normalization condition \eqref{Bloch_WP_normalization}.}.

The Bloch wave normalization condition~\eqref{F_normalization} implies $\int_{-\infty}^{+\infty} dx \, |\Psi(x,t)|^2 = \sum_{j = 0}^{\infty} \int_{-q/2}^{q/2} d\kappa \, \left| \phi^{(j)}(\kappa) \right|^2$. This means that normalizing the Bloch wave packet $\Psi(x,t)$ to unity is equivalent to requiring that the quasi-momentum amplitude $\phi^{(j)}(\kappa)$ satisfy
\begin{equation}
	\sum_{j = 0}^{\infty} \int_{-q/2}^{q/2} d\kappa \, \left| \phi^{(j)}(\kappa) \right|^2 = 1 \,.
\label{Bloch_WP_normalization}
\end{equation}

\section{Momentum positivity}
\label{sec:p>0}

We now turn to the question of momentum positivity for Bloch wave packets. Do there exist particle states in which a momentum measurement, performed at any instant, is guaranteed to yield a positive value? If such states existed -- analogous to those in free space -- they would correspond to particles with a well-defined momentum direction. However, as we show below, no Bloch wave packet can sustain a definite momentum direction throughout its propagation in periodic potentials with non-overlapping allowed energy bands.

We start by expanding the Bloch wave packet $\Psi$, given by Eq.~\eqref{Bloch_WP}, in terms of plane waves $\frac{1}{\sqrt{2 \pi}} e^{i k x}$, which are the eigenstates of the momentum operator $p = -i \hbar \frac{\partial}{\partial x}$,
\begin{equation}
	\Psi(x,t) = \int_{-\infty}^{+\infty} dk \, \varphi(k,t) \frac{e^{i k x}}{\sqrt{2 \pi}} \,.
\end{equation}
Here, $\varphi(k,t)$ plays the role of a time-dependent momentum wave function and is given by the inverse Fourier transform of $\Psi(x,t)$:
\begin{equation}
	\varphi(k,t) = \int_{-\infty}^{+\infty} dx \, \Psi(x,t) \frac{e^{-i k x}}{\sqrt{2 \pi}} \,.
\label{varphi_as_inverse_FT}
\end{equation}
Substituting Eq.~\eqref{Bloch_WP} into the last expression and calculating the integral over $x$, we find (see Appendix~\ref{app:momentum_wave_function})
\begin{equation}
	\varphi(k,t) = \sum_{j=0}^{\infty} \varphi^{(j)}(k,t) \,,
\label{varphi_as_sum}
\end{equation}
where
\begin{multline}
	\varphi^{(j)}(k,t) = \sqrt{2 \pi} \phi^{(j)}(k - n q) f_n^{(j)}(k - n q) \\ \times \left. e^{-i E^{(j)}(k - n q) t / \hbar} \right|_{n = [ k/q]}
\label{varphi_j_general}
\end{multline}
is the momentum wave function contribution from the $j^{\text{th}}$ allowed energy band, and $[\cdot]$ denotes rounding to the nearest integer.

If a momentum measurement is performed at time $t$ on a Bloch wave packet in the state $\Psi(x,t)$, the outcome $\hbar k$ is obtained with probability density  $|\varphi(k,t)|^2$. Therefore, the probability that the particle's momentum is positive at time $t$ is given by
\begin{equation}
	P_+(t) = \int_0^{\infty} dk \, |\varphi(k,t)|^2 \,.
\label{P_plus_def}
\end{equation}
In what follows, we show that no Bloch wave packets satisfy $P_+(t) = 1$ for all $t$, implying that a particle in a periodic potential cannot have a definite momentum direction at all times. Our argument proceeds in two stages: first, for wave packets composed of Bloch waves from a single allowed energy band, and then for those occupying multiple bands.

\subsection{Single-band wave packets}
\label{sec:single-band}

We first consider wave packets that are superpositions of Bloch waves from the same allowed energy band. That is, we take
\begin{equation}
	\phi^{(j)}(\kappa) = \phi(\kappa) \delta_{j,J}
\end{equation}
for some fixed $J$. The function $\phi(\kappa)$ is the quasi-momentum amplitude in the $J^{\text{th}}$ energy band and satisfies the normalization condition [c.f.~Eq.~\eqref{Bloch_WP_normalization}]
\begin{equation}
	\int_{-q/2}^{q/2} d\kappa \, |\phi(\kappa)|^2 = 1 \,.
\label{phi_normalization}
\end{equation}
It follows from Eqs.~\eqref{varphi_as_sum} and \eqref{varphi_j_general} that
\begin{equation}
	|\varphi(k,t)|^2 = \left. 2 \pi \left| \phi(k - n q) f_n^{(J)}(k - n q) \right|^2 \right|_{n = [ k/q]} \,.
\label{|varphi|^2}
\end{equation}
Then, substituting the last expression into Eq.~\eqref{P_plus_def} and carrying out straightforward manipulations (see Appendix~\ref{app:Lambda_derivation} for details), we obtain
\begin{equation}
	P_+ = \int_{-q/2}^{q/2} d\kappa \, \Lambda(\kappa) |\phi(\kappa)|^2
\label{P_plus_in_terms_of_Lambda}
\end{equation}
with
\begin{equation}
	\Lambda(\kappa) = 2 \pi \left| f_0^{(J)}(\kappa) \right|^2 \Theta(\kappa) + 2 \pi \sum_{n = 1}^{\infty} \left| f_n^{(J)}(\kappa) \right|^2 \,,
\label{Lambda_def}
\end{equation}
where $\Theta(\cdot)$ is the Heaviside step function.

Let us make the following observations. First, $P_+$ is independent of time, meaning that the probability of the particle's momentum being positive remains constant as it moves through the periodic potential. This situation is fully analogous to the free-particle scenario. Second, by comparing Eq.~\eqref{Lambda_def} with Eq.~\eqref{f_normalization}, we observe that $0 \le \Lambda(\kappa) \le 1$ for all $\kappa$. Combined with the normalization condition~\eqref{phi_normalization}, this implies that $0 \le P_+ \le 1$.

We now take this a step further and show that
\begin{equation}
	\Lambda(\kappa) < 1
\label{Lambda<1}
\end{equation}
for all $\kappa$. We prove this statement by contradiction. To this end, we assume that there exists an eigenvector $\{ f_n^{(J)}(\kappa) \}_{n=-\infty}^{+\infty}$, satisfying Eq.~\eqref{F_def} for $j=J$ and some fixed $\kappa$, such that $\Lambda(\kappa) = 1$, which, in view of Eq.~\eqref{Lambda_def}, is equivalent to
\begin{equation}
	\left| f_0^{(J)}(\kappa) \right|^2 \Theta(\kappa) + \sum_{n = 1}^{\infty} \left| f_n^{(J)}(\kappa) \right|^2  = \frac{1}{2 \pi} \,.
\label{assumption_1}
\end{equation}
Additionally, the eigenvector has to satisfy the normalization condition~\eqref{f_normalization}. Conditions~\eqref{f_normalization} and \eqref{assumption_1} can be simultaneously fulfilled only if
\begin{equation}
	f_n^{(J)}(\kappa) = 0 \quad \text{for} \quad n \le -1 \,.
\label{f_n=0_for_n<0}
\end{equation}
In view of Eq.~\eqref{f_n=0_for_n<0}, the eigenproblem~\eqref{F_def} reduces to
\begin{equation}
	\sum_{m=0}^{N-1} V_{n-m} f_m^{(J)}(\kappa) = 0 \quad \text{for} \quad -N \le n \le -1
\label{eigprob-1}
\end{equation}
and
\begin{multline}
	\sum_{m=0}^{\infty} \Bigg[ \left( \frac{\hbar^2 (\kappa + n q)^2}{2 \mu} - E^{(J)}(\kappa) \right) \delta_{m,n} \\ + V_{n-m} \Bigg] f_m^{(J)}(\kappa) =  0 \quad \text{for} \quad n \ge 0 \,.
\label{eigprob-2}
\end{multline}
It follows from Eq.~\eqref{eigprob-1} that
\begin{equation}
	f_n^{(J)}(\kappa) = 0 \quad \text{for} \quad 0 \le n \le N-1 \,.
\label{f_n=0_for_0<=n<N}
\end{equation}
Subsequently, from Eqs.~\eqref{eigprob-2} and \eqref{f_n=0_for_0<=n<N}, we get
\begin{equation}
	f_n^{(J)}(\kappa) = 0 \quad \text{for} \quad n \ge N \,.
\label{f_n=0_for_n>=N}
\end{equation}
(Details of the calculations leading to Eqs.~\eqref{f_n=0_for_0<=n<N} and \eqref{f_n=0_for_n>=N} are presented in Appendix~\ref{app:f_n=0}.) Combining Eqs.~\eqref{f_n=0_for_n<0}, \eqref{f_n=0_for_0<=n<N}, and \eqref{f_n=0_for_n>=N}, we find that $f_n^{(J)}(\kappa) = 0$ for all $n$, violating both the normalization condition~\eqref{f_normalization} and assumption~\eqref{assumption_1}. Hence, we arrive at a contradiction, proving that $\Lambda(\kappa)$ cannot equal unity and thus establishing that $\Lambda(\kappa) < 1$ for all $\kappa$.

It then follows from Eqs.~\eqref{P_plus_in_terms_of_Lambda} and \eqref{Lambda<1} that $P_+ < \int_{-q/2}^{q/2} d\kappa \, |\phi(\kappa)|^2$. Subsequently, using the normalization condition, Eq.~\eqref{phi_normalization}, we arrive at
\begin{equation}
	P_+ < 1 \,.
\label{main_result_1}
\end{equation}
This result implies that a single-band Bloch wave packet propagating through a periodic potential with a finite number of Fourier coefficients cannot have a well-defined momentum direction even instantaneously.

\subsection{Multiple-band wave packets}

We now consider normalized superpositions of Bloch waves from more than one allowed energy band. The substitution of Eqs.~\eqref{varphi_as_sum} and \eqref{varphi_j_general} into Eq.~\eqref{P_plus_def} yields
\begin{equation}
	P_+(t) = \overline{P} + \widetilde{P}(t) \,,
\label{P_plus_for_multiple_bands}
\end{equation}
with
\begin{equation}
	\overline{P} = \sum_j \int_0^{\infty} dk \, \left| \varphi^{(j)}(k,t) \right|^2
\end{equation}
and
\begin{equation}
	\widetilde{P}(t) = \sum_j \sum_{j' \not= j} \int_0^{\infty} dk \, \varphi^{(j)}(k,t)^* \varphi^{(j')}(k,t) \,.
\label{P_tilde}
\end{equation}
In the rest of this section, we do not explicitly state the summation limits to keep the notation uncluttered; however, they are clear, as both indices run from zero to infinity, with $j'$ skipping $j$.

The first term, $\overline{P}$, is a sum of (time-independent) single-band contributions of the type investigated in Sec.~\ref{sec:single-band}, each of which satisfies $\int_0^{\infty} dk \, \left| \varphi^{(j)}(k,t) \right|^2 < \int_{-q/2}^{q/2} d\kappa \, \left| \phi^{(j)}(\kappa) \right|^2$. Therefore, taking into account the normalization condition~\eqref{Bloch_WP_normalization}, we conclude that $\overline{P} < 1$.

The second term, $\widetilde{P}(t)$, is an interference term involving Bloch waves from different energy bands. Substituting Eq.~\eqref{varphi_j_general} into Eq.~\eqref{P_tilde} and following steps similar to those in Appendix~\ref{app:Lambda_derivation}, we obtain
\begin{multline}
	\widetilde{P}(t) = \sum_j \sum_{j' \not= j} \int_{-q/2}^{q/2} d\kappa \, \Lambda^{(j,j')}(\kappa) \phi^{(j)}(\kappa)^* \phi^{(j')}(\kappa) \\ \times \exp \left\{ \frac{i}{\hbar} \left[ E^{(j)}(\kappa) - E^{(j')}(\kappa) \right] t \right\}
\end{multline}
with
\begin{multline}
	\Lambda^{(j,j')}(\kappa) = 2 \pi f_0^{(j)}(\kappa)^* f_0^{(j')}(\kappa) \Theta(\kappa) \\ + 2 \pi \sum_{n=1}^{\infty} f_n^{(j)}(\kappa)^* f_n^{(j')}(\kappa) \,.
\end{multline}
Since $\Lambda^{(j,j')}(\kappa)^* = \Lambda^{(j',j)}(\kappa)$, the function $\widetilde{P}(t)$ is real-valued. Assuming, as before, that there is no overlap between allowed energy bands, we observe that $\widetilde{P}(t)$ is composed of sine and cosine functions of time with nonzero frequencies. Consequently, its time average is zero, implying that there are moments at which the function is non-positive: $\widetilde{P}(t) \le 0$ for some $t$.

Finally, from Eq.~\eqref{P_plus_for_multiple_bands} and the fact that $\overline{P} < 0$ and $\widetilde{P}(t) \le 0$ for some $t$, we conclude that
\begin{equation}
	P_+(t) < 1 \quad \text{for some } t \,.
\end{equation}
This result implies that no multiple-band Bloch wave packet can maintain a definite momentum direction at all times. While the probability of positive momentum may reach unity at certain instants, it must inevitably decrease, remaining below one on average.

\section{Cosine potential}
\label{sec:V=cos}

To examine how large $P_+$ can be depending on the details of the periodic potential, we analyze an example scenario where the potential is given by a cosine function,
\begin{equation}
	V(x) = A \cos(q x) \,,
	\label{V=cos}
\end{equation}
with an amplitude $A>0$. We consider the case where the wave packet $\Psi$ consists only of Bloch waves from the lowest allowed energy band, i.e.~$J=0$. The Fourier coefficients of this potential [c.f.~Eq.~\eqref{V(x)}] are $V_n = (A/2) \delta_{|n|,1}$, and the eigenproblem~\eqref{F_def} can be written (for $j=J=0$) as
\begin{multline}
	\sum_{m=-\infty}^{+\infty} \left[ (z+n)^2 \delta_{m,n} + \alpha \delta_{|n-m|,1} \right] f_m^{(0)}(z) \\ = \varepsilon(z) f_n^{(0)}(z) \,,
	\label{eigprob_for_cos_potential}
\end{multline}
where we have introduced the dimensionless quasi-momentum $z = \kappa/q$ (so that $-1/2 < z < 1/2$), dimensionless lowest band energy $\varepsilon(z) = 2 \mu E^{(0)}(\kappa) / (\hbar q)^2$, and dimensionless parameter
\begin{equation}
	\alpha = \frac{\mu A}{\hbar^2 q^2}
\end{equation}
quantifying the potential strength.

We solve Eq.~\eqref{eigprob_for_cos_potential} numerically by truncating the tridiagonal matrix on the left-hand side and diagonalizing it using the \texttt{eigh\_tridiagonal} function from the Python SciPy package. The diagonalization is performed on a grid of $z$ points ranging from $-1/2$ to $1/2$. The resulting eigenvectors, $\{ f_n^{(0)}(z) \}$, are then normalized for every value of $z$ in accordance with Eq.~\eqref{f_normalization}. The results presented below were obtained using truncated matrices of size $201 \times 201$ (corresponding to the index $m$ in Eq.~\eqref{eigprob_for_cos_potential} ranging from $-100$ to $100$), but we have verified that increasing the matrix size does not produce any noticeable changes in the results.

\begin{figure}[h]
	\centering
	\includegraphics[width=0.48\textwidth]{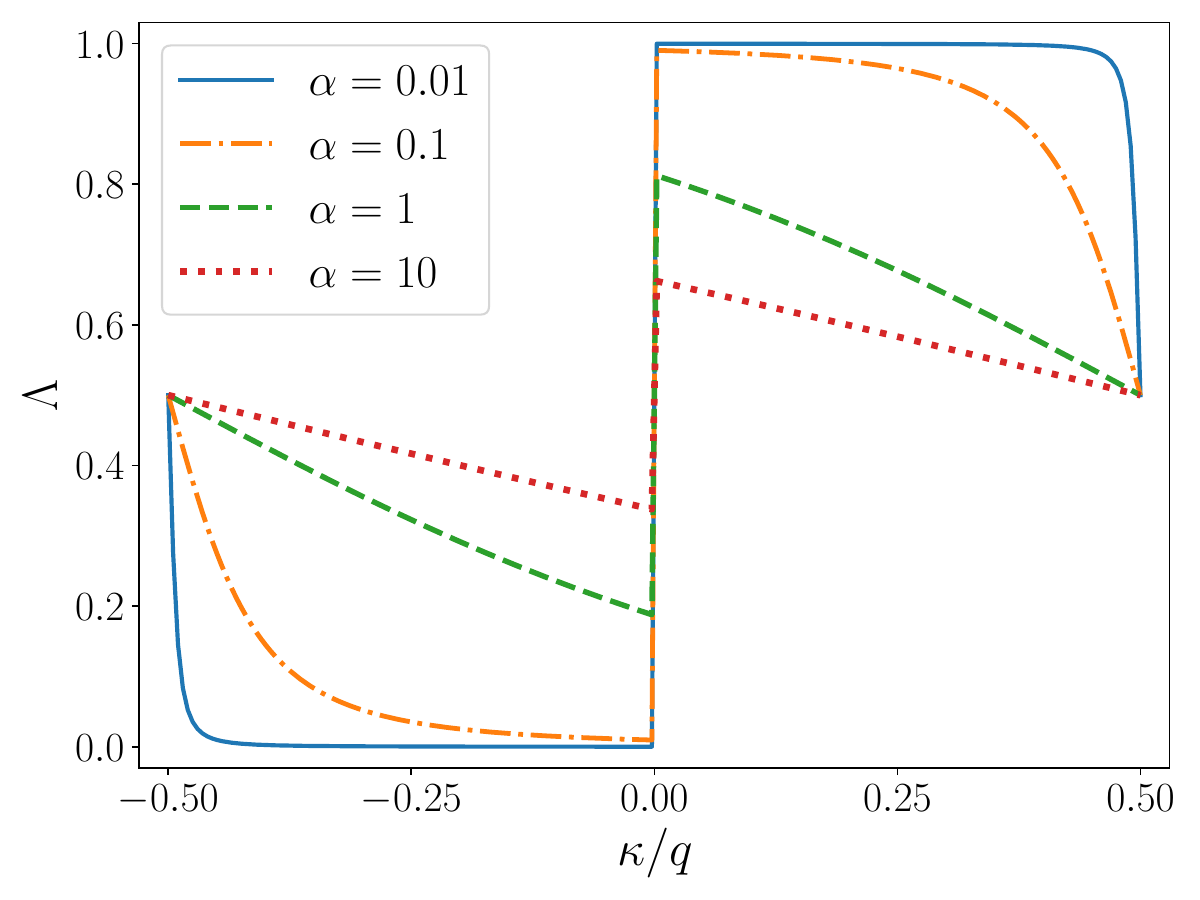}
	\caption{Dependence of $\Lambda$ on $z=\kappa/q$ for four different potential strengths, $\alpha = 0.01$, $0.1$, $1$, and $10$.}
	\label{fig:Lambda(z)}
\end{figure}

Equipped with the eigenvectors $\{ f_n^{(0)}(z) \}$, we compute the function $\Lambda(z)$, defined in Eq.~\eqref{Lambda_def}, for $J=0$, namely $\Lambda(z) = 2 \pi \left| f_0^{(0)}(z) \right|^2 \Theta(z) + 2 \pi \sum_{n = 1}^{\infty} \left| f_n^{(0)}(z) \right|^2$. Figure~\ref{fig:Lambda(z)} shows $\Lambda(z)$ for four different potential strength, $\alpha = 0.01$, $0.1$, $1$, and $10$. Having explored the function $\Lambda(z)$ for a wide range of $\alpha$ values, including the ones presented in Fig.~\ref{fig:Lambda(z)}, we conclude that
\begin{equation}
	\sup_z \Lambda = \Lambda(0^+) = 2 \pi \sum_{n = 0}^{\infty} \left| f_n^{(0)}(0) \right|^2 \,.
	\label{sup_Lambda}
\end{equation}
Taking into account the normalization condition, $2 \pi \sum_{n = -\infty}^{+\infty} \left| f_n^{(0)}(0) \right|^2 = 1$, and the fact that $f_{-n}^{(0)}(0) = f_n^{(0)}(0)$, which follows from Eq.~\eqref{eigprob_for_cos_potential}, we obtain that
\begin{equation}
	2 \pi \sum_{n = 0}^{\infty} \left| f_n^{(0)}(0) \right|^2 = \frac{1}{2} + \pi \left| f_0^{(0)}(0) \right|^2 \,.
	\label{f_n_symmetry}
\end{equation}
One can see from Eqs.~\eqref{P_plus_in_terms_of_Lambda} and \eqref{phi_normalization} that $\sup_{\phi} P_+ = \sup_z \Lambda$. Utilizing this fact, along with Eqs.~\eqref{sup_Lambda} and \eqref{f_n_symmetry}, we conclude that, for a given $\alpha$, the largest attainable value of $P_+$ is given by
\begin{equation}
	\sup P_+ = \frac{1}{2} + \pi \left| f_0^{(0)}(0) \right|^2 \,.
	\label{sup_P_plus}
\end{equation}

\begin{figure}[h]
	\centering
	\includegraphics[width=0.48\textwidth]{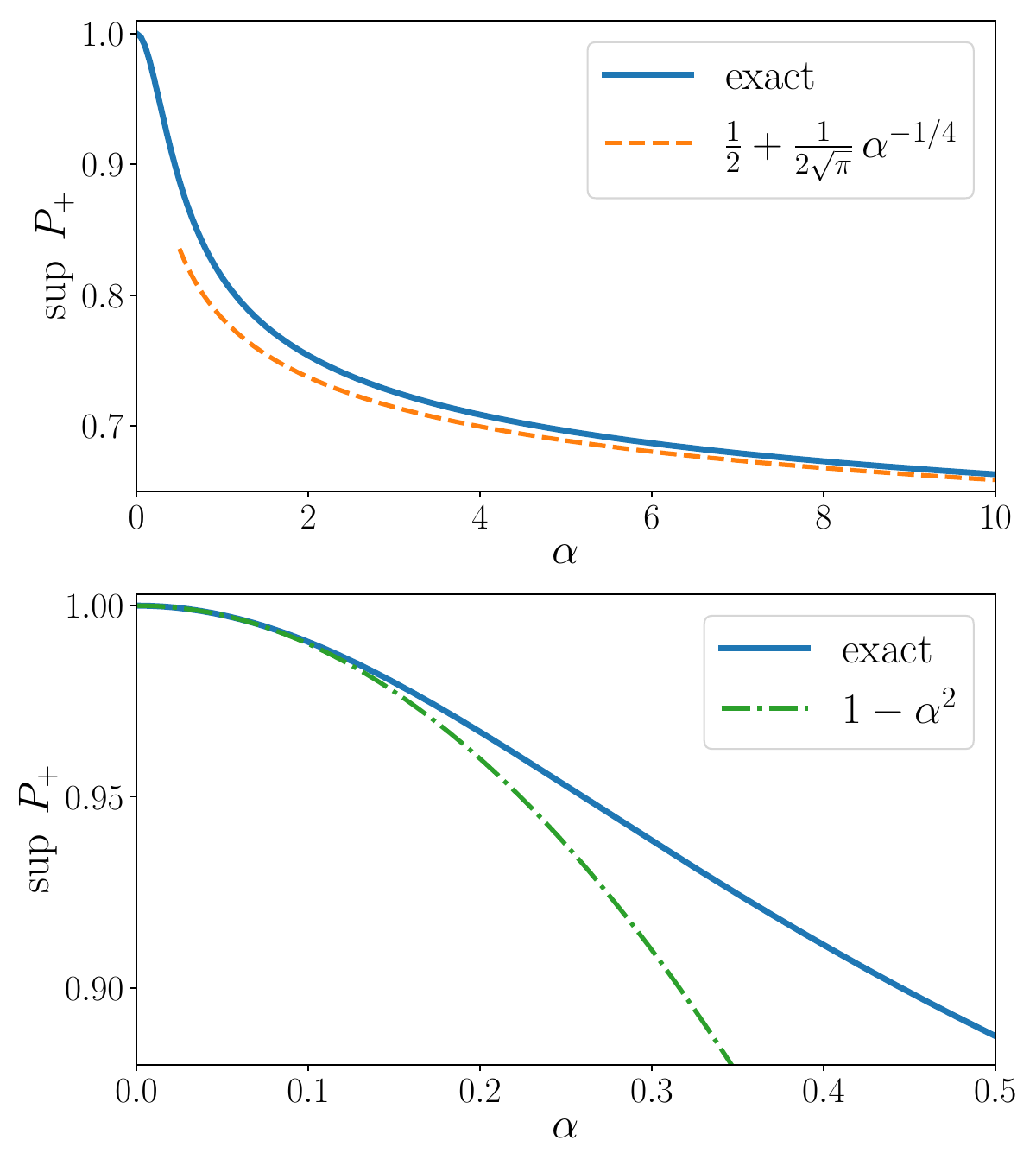}
	\caption{The largest attainable value of $P_+$ (blue solid curve) as a function of the potential strength $\alpha$. Additionally, the top panel shows the strong-potential asymptotics (orange dashed curve), while the bottom panel presents the weak-potential perturbative approximation (green dash-dotted curve).}
	\label{fig:sup_P}
\end{figure}

Figure~\ref{fig:sup_P} presents the dependence of $\sup P$ (blue solid curve) on the potential strength $\alpha$. The top panel displays this dependence over the broad interval $0 < \alpha < 10$, while the bottom panel provides a magnified view for $0 < \alpha < 0.5$. Additionally, the following two analytical approximations are included. The top panel shows the strong-potential approximation (orange dashed curve),
\begin{equation}
	\sup P_+ \simeq \frac{1}{2} + \frac{1}{2 \sqrt{\pi}} \alpha^{-1/4} \qquad (\alpha \gg 1) \,,
	\label{sup_P_for_alpha>>1}
\end{equation}
derived in Appendix~\ref{app:alpha>>1}. As we see from Eq.~\eqref{sup_P_for_alpha>>1}, $\sup P_+$ decays to $\frac{1}{2}$ as $\alpha$ tends to infinity, and the decay is relatively slow, $\sup P_+ - \frac{1}{2} \sim \alpha^{-1/4}$. The bottom panel presents the weak-potential approximation (green dash-dotted curve),
\begin{equation}
	\sup P_+ \simeq 1 - \alpha^2 \qquad (\alpha \ll 1) \,,
	\label{sup_P_for_alpha<<1}
\end{equation}
derived in Appendix~\ref{app:alpha<<1} by using the second-order perturbation theory.

The results presented in Fig.~\ref{fig:sup_P} support our claim that the momentum of a single-band wave packet propagating in a periodic potential can never have a definite direction at any instant. Moreover, we observe that the least upper bound on the probability of the particle having positive momentum decreases monotonically with increasing potential amplitude, approaching a saturation value of $1/2$ in the infinite-amplitude limit.

\section{Discussion and conclusion}
\label{sec:conclusion}

We examined the momentum of a one-dimensional quantum particle propagating through a periodic potential and investigated the probability $P_+$ of it being positive. We demonstrated that $P_+$ can never remain equal to one at all times, implying that the particle cannot propagate through the potential while maintaining a definite momentum direction. When the particle's state is a superposition of Bloch waves from a single allowed energy band, $P_+$ remains constant over time and is strictly less than one. However, if the state includes contributions from multiple energy bands, $P_+$ oscillates in time (around an average value smaller than one) and can, in principle, reach one at certain instants.

Our mathematical analysis involves three assumptions. First, the periodic potential is assumed to have a finite (though possibly very large) number of nonzero Fourier components, as given in Eq.~\eqref{V(x)}. Second, the allowed energy bands are assumed to be non-overlapping. Third, we consider only Bloch wave packets that do not include contributions from stationary states at the Brillouin zone boundary. Although we cannot, strictly speaking, rule out the possibility that relaxing some (or all) of these assumptions could allow for Bloch wave packets with a definite momentum direction, we consider this unlikely in any realistic physical system. Indeed, the first assumption is not physically restrictive, as any real-world periodic potential can be represented by a finite number of Fourier components. Examples of potentials $V(x)$ outside the scope of our analysis include functions that contain jump discontinuities or exhibit a fractal structure. As for the other two assumptions, allowing neighboring energy bands to touch at the center or edges of the Brillouin zone (the only way bands can overlap in one dimension~\cite{Cot71Floquet}), or including a finite number of states from the Brillouin zone boundary, would merely introduce a finite number of isolated Bloch waves. Such isolated states would contribute to the positive-momentum probability $P_+$, which is an integral quantity, only as a set of measure zero, and thus are unlikely to be sufficient to achieve $P_+ = 1$ at all times.

In general, the maximal value of $P_+$ depends on the specific form of the periodic potential and the energy bands accessible to the particle's wave packet. We investigated this dependence for a wave packet constrained to the lowest energy band of a cosine-shaped potential. In this scenario, we computed the least upper bound on $P_+$ as a function of the (dimensionless) potential amplitude $\alpha$ and found that $\sup P_+$ decreases monotonically with increasing $\alpha$, ultimately saturating at $1/2$ in the limit of an infinitely strong potential. In the weak-amplitude regime, the decay is quadratic [Eq.~\eqref{sup_P_for_alpha<<1}], whereas in the strong-amplitude regime, it follows an algebraic decay with an exponent of $-1/4$ [Eq.~\eqref{sup_P_for_alpha>>1}].

Our analysis of the cosine potential scenario also reveals that the wave packets maximizing $P_+$ are those composed of Bloch waves with small positive quasi-momenta -- indeed, the function $\Lambda(\kappa)$ in Fig.~\ref{fig:Lambda(z)} attains its supremum as $\kappa \to 0^+$. In this quasi-momentum region, the energy dispersion relation is approximately parabolic, resembling that of a free particle. This reinforces the well-known principle that, for a lowest-energy-band wave packet to propagate in a free-particle-like manner, it should be formed near the bottom of the band. However, it is crucial to emphasize that, regardless of how close a Bloch wave packet is to the bottom of the band, it will never possess a definite momentum direction -- unlike a free-particle wave packet, which can.

The findings of this paper provide valuable guidance for future explorations of the quantum backflow effect in mesoscopic systems. We identify two main strategies for studying quantum backflow in periodic potentials. The first involves selecting systems with shallow potentials, which, as our findings suggest, allow for wave packets with an {\it almost} well-defined momentum direction. By ensuring that the fraction of negative-momentum components remains small compared to the interference-induced negative probability flux throughout the experiment, one can access the backflow effect. In this context, quantum states confined to a single energy band appear preferable to multiple-band states, as the positive-momentum probability $P_+$ associated with the former remains constant during time evolution. The second strategy is to forgo the requirement of momentum positivity altogether and instead adopt a phase-space-based formulation of quantum backflow, similar to that examined in Refs.~\cite{MYDP21Experiment, BG21experiment}. However, given the challenges associated with the current phase-space approach, as highlighted in Ref.~\cite{BG21experiment}, we believe that further theoretical development is needed in this direction.

\begin{acknowledgments}
	The authors would like to thank Cosimo Gorini for valuable discussions.
\end{acknowledgments}

\appendix

\section{Derivation of Eq.~\eqref{F_normalization}}
\label{app:F_normalization}

This section provides a derivation of the normalization condition for Bloch waves, given by Eq.~\eqref{F_normalization}. Starting from Eq.~\eqref{Bloch_theorem}, we write
\begin{align*}
	\int_{-\infty}^{+\infty} &dx \, F^{(j)}(x, \kappa)^* F^{(j')}(x, \kappa') \\
	&= \int_{-\infty}^{+\infty} dx \, \left( e^{-i \kappa x} \sum_{m=-\infty}^{+\infty} f_m^{(j)}(\kappa)^* e^{-i m q x} \right) \\
	&\qquad\qquad\qquad \times \left( e^{i \kappa' x} \sum_{n=-\infty}^{+\infty} f_n^{(j')}(\kappa') e^{i n q x} \right) \\
	&= \sum_{m, n} f_m^{(j)}(\kappa)^* f_n^{(j')}(\kappa') \int_{-\infty}^{+\infty} dx \, e^{i [\kappa' - \kappa + (n-m) q] x} \\
	&= 2 \pi \sum_{m,n} f_m^{(j)}(\kappa)^* f_n^{(j')}(\kappa') \delta\big( \kappa - \kappa' + (m-n) q \big) \\
	&= 2 \pi \sum_{n = -\infty}^{+\infty} \sum_{s = -\infty}^{+\infty} f_{n+s}^{(j)}(\kappa)^* f_n^{(j')}(\kappa') \delta(\kappa - \kappa' + s q) \,.
\end{align*}
Since both $\kappa$ and $\kappa'$ are restricted to the first Brillouin zone, $\kappa, \kappa' \in (-q/2, q/2]$, their separation is smaller than the reciprocal lattice period, $|\kappa - \kappa'| < q$. Therefore, the delta-function $\delta(\kappa - \kappa' + s q)$ vanishes for all $s \not= 0$, yielding
\begin{multline*}
	\int_{-\infty}^{+\infty} dx \, F^{(j)}(x, \kappa)^* F^{(j')}(x, \kappa') \\
	= 2 \pi \delta(\kappa - \kappa') \sum_{n = -\infty}^{+\infty} f_n^{(j)}(\kappa)^* f_n^{(j')}(\kappa) \,.
\end{multline*}
For the case when $j' \not= j$, the eigenvectors $\{ f_n^{(j)}(\kappa) \}_{n=-\infty}^{+\infty}$ and $\{ f_n^{(j')}(\kappa) \}_{n=-\infty}^{+\infty}$ correspond to different energy bands (and so to distinct eigenvalues, $E^{(j)}(\kappa)$ and $E^{(j')}(\kappa)$, respectively), which implies their orthogonality. Hence,
\begin{multline*}
	\int_{-\infty}^{+\infty} dx \, F^{(j)}(x, \kappa)^* F^{(j')}(x, \kappa') \\
	= 2 \pi \delta(\kappa - \kappa') \delta_{j, j'} \sum_{n = -\infty}^{+\infty} \left| f_n^{(j)}(\kappa) \right|^2 \,.
\end{multline*}
Finally, making use of the normalization condition~\eqref{f_normalization}, we arrive at the desired result, given by Eq.~\eqref{F_normalization}.

\section{Derivation of Eqs.~\eqref{varphi_as_sum} and \eqref{varphi_j_general}}
\label{app:momentum_wave_function}

Here, we provided details of the calculation leading to Eqs.~\eqref{varphi_as_sum} and \eqref{varphi_j_general}. Substituting Eq.~\eqref{Bloch_WP} into Eq.~\eqref{varphi_as_inverse_FT}, we have
\begin{multline*}
	\varphi(k,t) = \sum_{j=0}^{\infty} \int_{-q/2}^{q/2} d\kappa \, \phi^{(j)}(\kappa) e^{-i E^{(j)}(\kappa) t / \hbar} \\
	\times \int_{-\infty}^{+\infty} dx \, F^{(j)}(x, \kappa) \frac{e^{-i k x}}{\sqrt{2 \pi}} \,.
\end{multline*}
Then, using Eq.~\eqref{Bloch_theorem}, we rewrite the last integral as
\begin{align*}
	\int_{-\infty}^{+\infty} &dx \, F^{(j)}(x, \kappa) \frac{e^{-i k x}}{\sqrt{2 \pi}} \\
	&= \frac{1}{\sqrt{2 \pi}} \sum_{n=-\infty}^{+\infty} f_n^{(j)}(\kappa) \int_{-\infty}^{+\infty} dx \, e^{i (\kappa + n q - k) x} \\
	&= \sqrt{2 \pi} \sum_{n=-\infty}^{+\infty} f_n^{(j)}(\kappa) \delta(\kappa + n q - k) \,.
\end{align*}
In view of this, the momentum wave function becomes
\begin{widetext}
\begin{align*}
	\varphi(k,t) &= \sqrt{2 \pi} \sum_{j=0}^{\infty} \sum_{n=-\infty}^{+\infty} \int_{-q/2}^{q/2} d\kappa \, \phi^{(j)}(\kappa) f_n^{(j)}(\kappa) e^{-i E^{(j)}(\kappa) t / \hbar} \delta \big( \kappa - (k - n q) \big) \\
	&= \sqrt{2 \pi} \sum_{j=0}^{\infty} \sum_{n=-\infty}^{+\infty} \phi^{(j)}(k - n q) f_n^{(j)}(k - n q) e^{-i E^{(j)}(k - n q) t / \hbar} \left\{
	\begin{array}{ll}
		1 \; &\text{if} \; |k - n q| < q/2 \\[0.1cm]
		0 \; &\text{otherwise}
	\end{array} \right. \\
	&= \left. \sqrt{2 \pi} \sum_{j=0}^{\infty} \phi^{(j)}(k - n q) f_n^{(j)}(k - n q) e^{-i E^{(j)}(k - n q) t / \hbar} \right|_{n = [k/q]} \,,
\end{align*}
\end{widetext}
where $[\cdot]$ denotes rounding to the nearest integer. The last expression for $\varphi$ coincides with the one given by Eqs.~\eqref{varphi_as_sum} and \eqref{varphi_j_general}.

\section{Derivation of Eqs.~\eqref{P_plus_in_terms_of_Lambda} and \eqref{Lambda_def}}
\label{app:Lambda_derivation}

Below we provide the calculation details leading to Eqs.~\eqref{P_plus_in_terms_of_Lambda} and \eqref{Lambda_def}. Substituting Eq.~\eqref{|varphi|^2} into Eq.~\eqref{P_plus_def} and representing the integral over the semi-infinite $k$ axis as a infinite sum of finite-interval integrals, we get
\begin{align*}
	P_+ &= \int_0^{q/2} dk \, 2 \pi \left| \phi(k) f_0^{(J)}(k) \right|^2 \\
	&\qquad + \sum_{n=1}^{\infty} \int_{nq - q/2}^{nq + q/2} dk \, 2 \pi \left| \phi(k-nq) f_n^{(J)}(k-nq) \right|^2 \\
	&= \int_{-q/2}^{q/2} dk \, \Theta(k) 2 \pi \left| \phi(k) f_0^{(J)}(k) \right|^2 \\
	&\qquad + \sum_{n=1}^{\infty} \int_{-q/2}^{q/2} dk \, 2 \pi \left| \phi(k) f_n^{(J)}(k) \right|^2 \\
	&= 2 \pi \int_{-q/2}^{q/2} dk \, \Bigg( \left| f_0^{(J)}(k) \right|^2 \Theta(k) \\
	&\qquad\qquad\qquad\qquad + \sum_{n=1}^{\infty} \left| f_n^{(J)}(k) \right|^2 \Bigg) |\phi(k)|^2 \,,
\end{align*}
which constitutes the desired expression.

\section{Derivation of Eqs.~\eqref{f_n=0_for_0<=n<N} and \eqref{f_n=0_for_n>=N}}
\label{app:f_n=0}

A straightforward way to demonstrate that Eq.~\eqref{f_n=0_for_0<=n<N} follows from Eq.~\eqref{eigprob-1} is to rewrite the latter in the matrix form:
\begin{equation*}
	\mat{V_{-N} & \ldots & V_{-3} & V_{-2} & V_{-1} \\ \vdots & \ddots & \vdots & \vdots & \vdots \\[0.1cm] 0 & \ldots & V_{-N} & V_{-N+1} & V_{-N+2} \\[0.2cm] 0 & \ldots & 0 & V_{-N} & V_{-N+1} \\[0.2cm] 0 & \ldots & 0 & 0 & V_{-N}} \mat{f_{N-1}^{(J)} \\ \vdots \\[0.1cm] f_2^{(J)} \\[0.15cm] f_1^{(J)} \\[0.1cm] f_0^{(J)}} = 0 \,.
\end{equation*}
Hereinafter, for the sake of notational brevity, we have omitted the $\kappa$ dependence of the coefficients $f_n^{(J)}(\kappa)$. The matrix in the left-hand side of this last equation is nonsingular, as its determinant is $(V_{-N})^N \not= 0$. Therefore, the equation is satisfied only by the zero vector, implying Eq.~\eqref{f_n=0_for_0<=n<N}.

The proof of Eq.~\eqref{f_n=0_for_n>=N} proceeds by induction. To this end, we observe that if $f_0^{(J)} = f_1^{(J)} = \ldots = f_{\ell - 1}^{(J)} = 0$, for any $\ell \ge N$, then Eq.~\eqref{eigprob-2} yields $f_{\ell}^{(J)} = 0$. Indeed, writing  Eq.~\eqref{eigprob-2} for $n = \ell - N \ge 0$,
\begin{multline*}
	\sum_{m=0}^{\infty} \Bigg[ \left( \frac{\hbar^2 \big( \kappa + (\ell - N) q \big)^2}{2 \mu} - E^{(J)} \right) \delta_{m,\ell-N} \\ + V_{\ell - N - m} \Bigg] f_m^{(J)} = 0 \,,
\end{multline*}
taking into account that $f_0^{(J)} = \ldots = f_{\ell - 1}^{(J)} = 0$,
\begin{equation*}
	\sum_{m=\ell}^{\infty} V_{\ell - N - m} f_m^{(J)} = 0 \,,
\end{equation*}
and, recognizing the fact that $V_n = 0$ for all $n < -N$, we arrive at
\begin{equation*}
	V_{-N} f_{\ell}^{(J)} = 0 \,.
\end{equation*}
Since $V_{-N} \not= 0$, this implies $f_{\ell}^{(J)} = 0$. Thus, by the principle of induction, we obtain Eq.~\eqref{f_n=0_for_n>=N}.

\section{Derivation of Eq.~\eqref{sup_P_for_alpha>>1}}
\label{app:alpha>>1}

Here, we derive the large-$\alpha$ asymptotic approximation for $\sup P_+$, as given by Eq.~\eqref{sup_P_for_alpha>>1}. We begin by noticing that the zero-quasi-momentum Bloch wave $F^{(0)}(x,0)$ is an $L$-periodic function [see Eq.~\eqref{Bloch_theorem}], which, over the one-period interval $0 < x < L$, can be approximated by ground state of the harmonic expansion of $V(x)$ around $x = L/2$. Thus, using, in the vicinity of $x = L/2$, the quadratic expansion
\begin{align*}
	V(x) &\simeq V(\tfrac{L}{2}) + \frac{V''(\tfrac{L}{2})}{2} \left( x - \tfrac{L}{2} \right)^2 \\
	&= -A + \frac{\mu \omega^2}{2} \left( x - \tfrac{L}{2} \right)^2
\end{align*}
with
\begin{equation*}
	\omega = \sqrt{\frac{A q^2}{\mu}} = \frac{\hbar q^2}{\mu} \sqrt{\alpha} \,,
\end{equation*}
we write, for $0 < x < L$, that
\begin{align*}
	F^{(0)}(x,0) &\simeq  C \exp \left[ -\frac{\mu \omega}{2 \hbar} \left( x - \frac{L}{2} \right)^2 \right] \\
	&= C \exp \left[ -\frac{\sqrt{\alpha}}{2} q^2 \left( x - \tfrac{L}{2} \right)^2 \right] \,,
\end{align*}
where $C>0$ is a normalization constant. The latter is determined from the normalization condition
\begin{equation*}
	\int_0^L dx \, \left| F^{(0)}(x,0) \right|^2 = \frac{1}{q} \,.
\end{equation*}
The validity of this normalization condition follows straightforwardly from Eqs.~\eqref{Bloch_theorem} and \eqref{f_normalization}:
\begin{align*}
	\int_0^L dx \, &\left| F^{(0)}(x,0) \right|^2 = \int_0^L dx \, \sum_{m=-\infty}^{+\infty} f_m^{(0)}(0)^* e^{-i m q x} \\
	&\qquad\qquad\qquad\qquad\qquad \times \sum_{n=-\infty}^{+\infty} f_n^{(0)}(0) e^{i n q x} \\
	&= \sum_{m, n} f_m^{(0)}(0)^* f_n^{(0)}(0) \int_0^L dx \, e^{i (n-m) q x} \\
	&= \sum_{m, n} f_m^{(0)}(0)^* f_n^{(0)}(0) L \delta_{mn} \\
	&= L \sum_{n=-\infty}^{\infty} \left| f_n^{(0)}(0) \right|^2 = L \frac{1}{2 \pi} = \frac{1}{q} \,.
\end{align*}
Hence,
\begin{equation*}
	C^2 \int_0^L dx \, \exp \left[ -\sqrt{\alpha} q^2 \left( x - \frac{L}{2} \right)^2 \right] = \frac{1}{q} \,.
\end{equation*}
For large $\alpha$, we extend the integration interval to the entire line, thus obtaining $C^2 \sqrt{ \pi / (\sqrt{\alpha} q^2) } \simeq 1/q$, or
\begin{equation*}
	C \simeq \frac{\alpha^{1/8}}{\pi^{1/4}} \,.
\end{equation*}
Therefore, we arrive at the following (normalized) large-$\alpha$ approximation for the lowest-allowed-band zero-quasi-momentum Bloch wave on the interval $0 < x < L$:
\begin{equation*}
	F^{(0)}(x,0) \simeq \frac{\alpha^{1/8}}{\pi^{1/4}} \exp \left[ -\frac{\sqrt{\alpha}}{2} q^2 \left( x - \frac{L}{2} \right)^2 \right] \,.
\end{equation*}

We now calculate the $n=0$ Fourier coefficient of $F^{(0)}(x,0)$ [see Eq.~\eqref{Bloch_theorem}]:
\begin{align*}
	f_0^{(0)}(0) &= \frac{1}{L} \int_0^L dx \, F^{(0)}(x, 0) \\
	&\simeq \frac{\alpha^{1/8}}{\pi^{1/4}} \frac{1}{L} \int_{-\infty}^{+\infty} dx \, \exp \left[ -\frac{\sqrt{\alpha}}{2} q^2 \left( x - \frac{L}{2} \right)^2 \right] \\
	&= \frac{\alpha^{1/8}}{\pi^{1/4}} \frac{1}{L} \sqrt{\frac{2 \pi}{\sqrt{\alpha} q^2}} = \frac{1}{\sqrt{2} \pi^{3/4}} \alpha^{-1/8} \,.
\end{align*}
Finally, substituting this expression for $f_0^{(0)}(0)$ into Eq.~\eqref{sup_P_plus}, we obtain the sought result, Eq.~\eqref{sup_P_for_alpha>>1}.

\section{Derivation of Eq.~\eqref{sup_P_for_alpha<<1}}
\label{app:alpha<<1}

Below we derive Eq.~\eqref{sup_P_for_alpha<<1}, approximating $\sup P_+$ in the weak-amplitude limit. Our strategy is to find a small-$\alpha$ perturbative expression for the Fourier coefficient $f_0^{(0)}(0)$ and then use Eq.~\eqref{sup_P_plus} to obtain $\sup P_+$.

For convenience, we carry out the calculation using scaled Fourier coefficients
\begin{equation*}
	g_n^{(j)} \equiv \sqrt{2 \pi} f_n^{(j)}(0) \,,
\end{equation*}
in terms of which the normalization condition~\eqref{f_normalization} reads $\sum_{n=-\infty}^{+\infty} \left| g_n^{(j)} \right|^2 = 1$. Since we assume throughout this paper that there is no overlap between different allowed energy bands, the vectors $g^{(j)}$ and $g^{(j')}$ with $j' \not= j$ are mutually orthogonal. Combined with the normalization condition, this leads to the orthonormality condition
\begin{equation*}
	\sum_{n=-\infty}^{+\infty} \left[ g_n^{(j)} \right]^* g_n^{(j')} = \delta_{j,j'} \,.
\end{equation*}

Written for vectors $g^{(j)}$, the eigenproblem~\eqref{F_def} takes the form
\begin{equation*}
	(H + U) g^{(j)} = \varepsilon^{(j)} g^{(j)} \,,
\end{equation*}
where $H$ is the unperturbed (diagonal) matrix with elements
\begin{equation*}
	H_{m n} = n^2 \delta_{m n} \,,
\end{equation*}
$U$ is the perturbation matrix with elements
\begin{equation*}
	U_{m n} = \alpha \delta_{|m-n|, 1} \,,
\end{equation*}
and $\varepsilon^{(j)} = 2 \mu E^{(j)}(0) / (\hbar q)^2$ is the corresponding dimensionless eigenvalue. To the second order in the perturbation parameter $\alpha$, we write
\begin{equation*}
	g^{(j)} \simeq g^{(j)} |_0 + g^{(j)} |_1 + g^{(j)} |_2 \,.
\end{equation*}
where the unperturbed eigenvector $g^{(j)} |_0$ has elements $g^{(j)}_n |_0 = \delta_{j,n}$. The first and second order corrections, $g^{(j)} |_1$ and $g^{(j)} |_2$ respectively, are given by \cite{Landau_Lifshitz}
\begin{widetext}
\begin{equation*}
	g^{(j)} |_1 = \sum_{m \not= j} \frac{U_{m j}}{j^2 - m^2} g^{(m)} |_0
\end{equation*}
and
\begin{equation*}
	g^{(j)} |_2 = \sum_{m \not= j} \sum_{n \not= j} \frac{U_{m n} U_{n j}}{(j^2 - m^2) (j^2 - n^2)} g^{(m)} |_0 - \sum_{m \not= j} \frac{U_{j j} U_{m j}}{(j^2 - m^2)^2} g^{(m)} |_0 - \frac{g^{(j)} |_0}{2} \sum_{m \not= j} \frac{|U_{m j}|^2}{(j^2 - m^2)^2} \,.
\end{equation*}
Evaluating these expressions for $j=0$, we get
\begin{equation*}
	g_0^{(0)} |_1 = \sum_{m \not= 0} \frac{U_{m 0}}{-m^2} g_0^{(m)} |_0 = \sum_{m \not= 0} \frac{\alpha \delta_{|m|,1}}{-m^2} \delta_{m 0} = 0
\end{equation*}
and
\begin{align*}
	g_0^{(0)} |_2 &= \sum_{m \not= 0} \sum_{n \not= 0} \frac{U_{m n} U_{n 0}}{(-m^2) (-n^2)} g_0^{(m)} |_0 - \sum_{m \not= 0} \frac{U_{0 0} U_{m 0}}{(-m^2)^2} g_0^{(m)} |_0 - \frac{g_0^{(0)} |_0}{2} \sum_{m \not= 0} \frac{|U_{m 0}|^2}{(-m^2)^2} \\[0.2cm]
	&= \sum_{m \not= 0} \sum_{n \not= 0} \frac{\alpha \delta_{|m-n|, 1} \alpha \delta_{|n|,1}}{(-m^2) (-n^2)} \delta_{m 0} - \sum_{m \not= 0} \frac{\alpha \delta_{0,1} \alpha \delta_{|m|,1}}{(-m^2)^2} \delta_{m 0} - \frac{1}{2} \sum_{m \not= 0} \frac{\left( \alpha \delta_{|m|,1} \right)^2}{(-m^2)^2} \\[0.2cm] &= - \frac{1}{2} \sum_{m \not= 0} \frac{\left( \alpha \delta_{|m|,1} \right)^2}{(-m^2)^2} = -\frac{1}{2} \left( \frac{\alpha^2}{(-1^2)^2} + \frac{\alpha^2}{(-(-1)^2)^2} \right) = -\alpha^2 \,.
\end{align*}
\end{widetext}
Therefore, up to the terms of order $\alpha^2$, we have $g_0^{(0)} \simeq 1 - \alpha^2$, or, equivalently,
\begin{equation*}
	f_0^{(0)}(0) \simeq \frac{1 - \alpha^2}{\sqrt{2 \pi}} \,.
\end{equation*}
Finally, substituting this expression into Eq.~\eqref{sup_P_plus} and discarding terms of order higher than $\alpha^2$, we obtain the sought result.

%

\end{document}